\documentclass{article}
\usepackage{dcase2018,amsmath,graphicx,url,times,booktabs, tabularx}
\usepackage[caption=false, font=small]{subfig}
\usepackage{amsfonts,array}
\usepackage{bbm}
\usepackage{color}
\title{Unsupervised adversarial domain adaptation for acoustic scene classification}
\name{Shayan Gharib$^{1*}$, Konstantinos Drossos$^{1}\sthanks{Equally contributing authors}$, Emre \c{C}akir$^{1}$, Dmitriy Serdyuk$^{2}$, and Tuomas Virtanen$^{1}$}
\address{$^{1}$Audio Research Group, Lab. of Signal Processing,\\Tampere University of Technology, Tampere, Finland\\
$^{2}$Montreal Institute for Learning Algorithms, Montreal, Canada}
\begin{document}
\ninept\maketitle\begin{sloppy}
\begin{abstract}
A general problem in acoustic scene classification task is the mismatched conditions between training and testing data, which significantly reduces the performance of the developed methods on classification accuracy. As a countermeasure, we present the first method of unsupervised adversarial domain adaptation for acoustic scene classification. We employ a model pre-trained on data from one set of conditions and by using data from other set of conditions, we adapt the model in order that its output cannot be used for classifying the set of conditions that input data belong to. We use a freely available dataset from the DCASE 2018 challenge Task 1, subtask B, that contains data from mismatched recording devices. We consider the scenario where the annotations are available for the data recorded from one device, but not for the rest. Our results show that with our model agnostic method we can achieve $\sim 10\%$ increase at the accuracy on an unseen and unlabeled dataset, while keeping almost the same performance on the labeled dataset. 
\end{abstract}
\vspace{-8pt}
\section{Introduction}\label{sec:intro}
\vspace{-6pt}
The task of acoustic scene classification is to assign to a sound segment the acoustic scene that it belongs to, e.g. office, park, tram, etc. Recently proposed methods for acoustic scene classification (ASC) are based on deep neural networks (DNNs)\cite{valenti2016dcase, han2017convolutional}. They usually employ convolutional neural networks (CNNs) to extract discriminative features from the used data, then using these features as an input to a classifier for classifying the acoustic scene~\cite{han2017,weiping2017,lehner2017,park2017}. In a realistic scenario, a method for ASC will be used to classify data emerging from a variety of different domains (i.e. acoustic conditions, acoustic channels) from the data used for optimizing that particular method. The mismatched domains introduce the dataset bias (or domain shift) phenomenon~\cite{gretton:2009:chapter, ganin:icml:2015, tzeng:cvpr:2017}, which results in a degradation of the performance of the method.

A typical countermeasure to this phenomenon is the fine tuning of the method using annotated data from different acoustic conditions. For example, one can retrain a method given a newly collected dataset. But, the annotation of audio data is a tedious process and it is more likely for one to have audio data but not having their annotations. To leverage knowledge from new and unlabeled data, one can use domain adaptation processes. Domain adaptation is a subspace alignment problem, where the goal is the alignment of the latent representations of the data coming from different domains~\cite{ganin:icml:2015,pei:aaai:2018,fernando:ccv:2013,alam:acl:2018}. The impact of the domain adaptation is greater when none or few annotations (labels) exist for data from the different domains. These processes are referred as unsupervised and semi-supervised, respectively, domain adaptation. 

Before the emergence of adversarial training, different approaches had been employed to cope with the problem of covariate shift across domains, e.g. kernel mean matching (KMM)~\cite{gretton2009covariate, miao2012cross} and autoencoder scheme based approaches~\cite{deng2014autoencoder, glorot2011domain, chen2012marginalized}. One of the first works in adversarial domain adaptation with deep neural networks is~\cite{ganin2016domain} where the classification and the alignment of the latent representations can occur at the same time. The alignment is performed by using the reverse gradient of the domain classification to optimize the parameters that produce the latent representation for the classification. A similar concept has been adopted in many subsequent works, e.g.~\cite{pei2018multi}. Later, an adversarial domain adaptation approach is presented in~\cite{tzeng:cvpr:2017}, where the training procedure of classification and adaptation are not happening simultaneously. The first step obtains a non-adapted model and, in a second step, this model is adapted. This increases the performance of adaptation, compared to the previous existing methods. Another method used in~\cite{bousmalis2016domain} implements classification and reconstruction by employing three different feature extractors and one shared encoder. One of the feature extractors is shared between the domains, while the other two are domain exclusive. The classifier predicts the labels based on the shared features between domains. In addition, there is an adversarial objective function for shared features to help the adaptation by increasing the similarity of extracted features across domains. Another recent work~\cite{rozantsev2018beyond} presents two models for source and target while regularizing their parameters by sharing a loss between each layer, targeting to mitigate the existing disparity between source and target distributions. The above methods evaluate the domain adaptation in the context of natural language processing, sentiment classification, and image classification. There are no previous studies in the context of acoustic scene classification.

Driven by the above, in this paper we present the first approach for unsupervised domain adaptation for acoustic scene classification. We investigate the unsupervised domain adaptation scenario, i.e. the acoustic scene labels of the new data are not known during the adaptation part. We use the data from the DCASE 2018 Task 1, subtask B, which consist of recordings from mismatched recording devices~\cite{dcase2018:arxiv}. We consider the difference in the acoustic channel, imposed by the different recording devices, as the domains. To mitigate this difference, we introduce a model agnostic process where we encourage the model to match the distributions of the learned representations of the data coming from the annotated (source domain) and the non-annotated (target domain) sets. The contributions of this paper are the following: 
\begin{enumerate}
\item We follow a recently proposed general framework for adversarial domain adaptation~\cite{tzeng:cvpr:2017} and we alter it by introducing extra learning signals during the adaptation process;
\item We present the first application of deep neural network based, unsupervised domain adaptation for acoustic scene classification showing the effect of leveraging unlabeled data for acoustic scene classification, through unsupervised domain adaptation.
\end{enumerate}
\begin{figure*}[!ht]
\centering
\includegraphics[width=.9\textwidth]{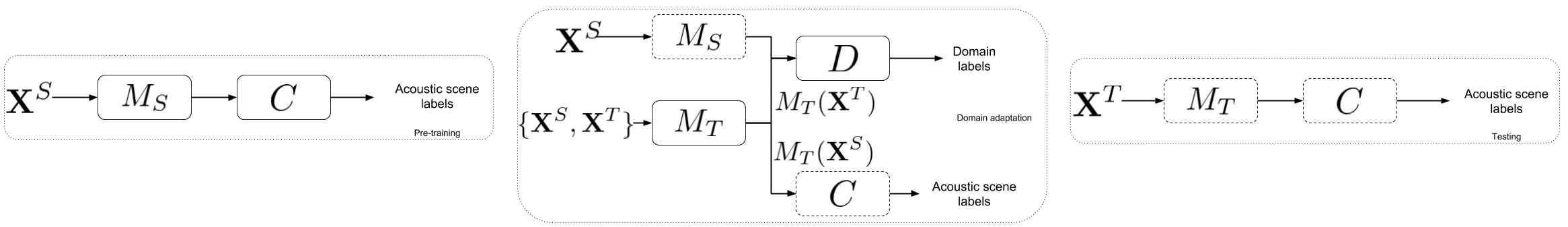}
\caption{Illustration of the three steps of the domain adaptation method: pre-training; adversarial domain adaptation; and testing. Solid lines indicate the models that are optimized in the corresponding steps, and dashed lines indicate the models that are not optimized.}
\label{fig:method}
\end{figure*}
The rest of the paper is organized as follows. In Section~\ref{sec:method} we explain the proposed method, and in Section~\ref{sec:evaluation} we present the evaluation procedure that we followed, including the presentation of the dataset used and the models implemented, and the details of the training and testing procedures. The obtained results are reported and discussed in Section~\ref{sec:results}, followed by the conclusions and proposals for future work in Section~\ref{sec:conclusions}.
\vspace{-8pt}
\section{Proposed domain adaptation method}\label{sec:method}
\vspace{-6pt}
The data from the source domain are classified according to a specific set of labeled acoustic scenes. For example, in our study, these acoustic scenes are \textit{airport}, \textit{bus}, \textit{airport}, \textit{metro}, \textit{metro station}, \textit{park}, \textit{public square}, \textit{shopping mall}, \textit{street pedestrian}, \textit{street traffic}, and \textit{tram}.  The goal is to assign the target domain data to this same set of labels. The source and target domains data are time-frequency representation of audio (e.g. log mel-band energies). We follow the general framework for adversarial domain adaptation in~\cite{tzeng:cvpr:2017} and choose not to tie the parameters of the source and adapted (target) models. Our models are neural networks that are used to extract the discriminative latent representation of the input data. This representation is used for the label classification by the classifier. The source model is the model optimized with the source data and the target model is the one adapted to the target data. Our presented method is independent of the architecture of the utilized model and concerns the adaptation of a model optimized on the source domain, to the target domain.  For this reason, in this section, we present the method for the domain adaptation, and in Section~\ref{sec:evaluation} we present the specific models employed.

Having annotated (i.e. with reference labels) data from the source domain, $\mathbf{X}^{S}=\{\mathbf{X}^{S}_{1},\mathbf{X}^{S}_{2},\ldots,\mathbf{X}^{S}_{N_{S}}\}$, and the non-annotated target domain data, $\mathbf{X}^{T}=\{\mathbf{X}^{T}_{1},\mathbf{X}^{T}_{2},\ldots,\mathbf{X}^{T}_{N_{T}}\}$, the goal is to regularize a model $M$ to produce feature mappings of the source domain, $M(\mathbf{X}^{S})$, and of the target domain, $M(\mathbf{X}^{T})$, that exhibit the same distribution. Then a classifier, trained on $M(\mathbf{X}^{S})$, can be used in order to classify $M(\mathbf{X}^{T})$. For this process, we employ three steps. At the first step, we pre-train the model $M$ and the classifier $C$ using dataset $\mathbf{X}^{S}$. Then, at the second step, we use  adversarial training (as in generative adversarial network (GAN) \cite{NIPS2014_5423}) to match the distributions of $M(\mathbf{X}^{S})$ and $M(\mathbf{X}^{T})$. Finally, at the third step, we test the performance of the classifier on the $M(\mathbf{X}^{T})$. All the three steps of the process are schematically illustrated at Figure~\ref{fig:method}. 

We differ from the original proposal of the general framework for adversarial domain adaptation in~\cite{tzeng:cvpr:2017} by utilizing the label classifier $C$ also during the adaptation step. Also, we differ from proposals with gradient reversing, e.g.~\cite{ganin:icml:2015}, because the classifier $C$ is not the domain classifier but the label one. We experimentally found that, for our task, the original setup (i.e. without $C$ in the adaptation step) cannot work. In this setup, the adapted model was exhibiting worse label classification performance to both the target and the source domains, compared to the non-adapted one. Observing the adaptation process, we hypothesized that the learning signals used in~\cite{tzeng:cvpr:2017} were not able to drive the model $M$ to produce feature mappings that can be used for later target domain classification from $C$. Thus, we utilize the $C$ in order to provide an additional learning signal during the adaptation process. This results in more stable domain adaptation process and the adapted model exhibits increased performance at the target domain, compared to the non-adapted one.

We start by having the data from the source domain, $\mathbf{X}^{S}$, and their corresponding one-hot-encoded labels for the acoustic scene, $\mathbf{Y}^{S} = \{\mathbf{y}^{S}_{1}, \mathbf{y}^{S}_{2}, \ldots, \mathbf{y}^{S}_{N_{S}}\}$. The first goal is to obtain a model and a classifier that are able to classify the source data (i.e. label and not domain classification). To this end, we utilize the $\mathbf{X}^{S}$ and $\mathbf{y}^{S}$ to train our source domain model, $M_{S}$, and pretrain the classifier $C$, by minimizing the loss
\begin{equation}
\mathcal{L}_{S} = -\sum\limits_{n=1}^{N_{S}}\mathbf{y}^{S}_{n}\log(C(M_{S}(\mathbf{X}^{S}_{n})))\text{,}
\end{equation}

At the second step, we target to obtain a model that can produce mappings of the data from the source and target domains, that they are as close as possible in terms of their distribution. Since we do not have the labels of the target domain, we can only leverage knowledge from the source data and their labels, and from the data of the target domain. Adopting the approach in~\cite{tzeng:cvpr:2017}, we use the adversarial training to match the distributions of $M(\mathbf{X}^{S})$ and $M(\mathbf{X}^{T})$. Specifically, we use an additional, target domain model, $M_{T}$, having the same architecture and amount of parameters as $M_{S}$. We do not use any constraints between $M_{S}$ and $M_{T}$ (e.g. parameters sharing/coupling between $M_{S}$ and $M_{T}$), but we initialize the parameters of $M_{T}$ with the ones from $M_{S}$. Additionally, we use a domain discriminator $D$ that will be optimized to identify if its input is coming from the distribution of the source or the target domain (hence, its output is an indication if its input was or not from the source domain). 

We jointly optimize the $M_{T}$ and $D$ in order to enforce the distribution of the $M_{T}(\mathbf{X}^{T})$ to be as close as possible to the distribution of the $M_{S}(\mathbf{X}^{S})$. In the GAN terminology, one can think the $M_{T}$ as the generator, the $M_{S}$ as the real examples, and the $D$ as the discriminator. The output of the generator and real examples are given as an input to the discriminator, and the latter is optimized to identify which is real and which is coming from the generator. At the same time, the generator is optimized to fool the discriminator in believing that the output of the generator is also a real example. In our method, $M_{S}(\mathbf{X}^{S})$ (real examples) and $M_{T}(\mathbf{X}^{T})$ (generator output) are given as an input to the discriminator $D$. The latter is optimized to identify if its input is $M_{S}(\mathbf{X}^{S})$ or $M_{T}(\mathbf{X}^{T})$. At the same time, we optimize $M_{T}$ in order to fool $D$ that $M_{T}(\mathbf{X}^{T})$ is $M_{S}(\mathbf{X}_{S})$. We minimize the losses
\begin{flalign}
\mathcal{L}_{D} = &-\sum_{n=1}^{N_{S}}(\log D(M_S (\mathbf{X}^{S}_{n}) + \log (1 - D(M_{T}(\mathbf{X}^{T}_{n})))\text{ and}\\
\mathcal{L}_{M_{T}} = &-\sum\limits_{n=1}^{N_{S}}(\log D(M_{T}(\mathbf{X}^{T}_{n})+ \mathbf{y}_{S_{n}}\log(C(M_{T}(\mathbf{X}^{S}_{n})))\text{.}
\end{flalign}
\noindent
$\mathcal{L}_{D}$ is minimized w.r.t $D$ and the $\mathcal{L}_{M_{T}}$ w.r.t. $M_{T}$. In the case where $N_{T}<N_{S}$ or $N_{T}>N_{S}$, then $\mathbf{X}^{T}$ will be either oversampled or undersampled, respectively. The minimization of $\mathcal{L}_{D}$ and $\mathcal{L}_{M_{T}}$ can be performed jointly or in an alternating way, e.g. do an update of $D$ towards minimizing $\mathcal{L}_{D}$, then update $M_{T}$ towards minimizing $\mathcal{L}_{M_{T}}$, and repeat until some criterion is met. The actual implementation of the minimization process is tied to the employed models of $M_{S}$, $M_{T}$, $D$, and $C$, and the dynamics of the training process. Our followed procedure is presented in Section~\ref{sec:evaluation}. Finally, we use $M_{T}$ with $C$ in order to classify the $\mathbf{X}^{T}$. 
\vspace{-8pt}
\section{Evaluation}\label{sec:evaluation}
\vspace{-6pt}
To assess the performance of our method we focus on the task of acoustic scene classification. We employ a freely available dataset that provides audio data recorded with mismatched recording devices. For model $M$, we employ two different models; one that achieved the first place in an acoustic scene classification contest\footnote{\url{https://www.kaggle.com/c/acoustic-scene-2018}} and a second that is the published baseline model for the Task 1, subtask B, of the DCASE challenge 2018~\cite{dcase2018:arxiv}. For the rest of the paper, we will refer to the former model as the Kaggle model and to the latter as the DCASE model. All hyper-parameters reported in this section are after a grid search process, using the validation data. All models are implemented using the freely available PyTorch framework\footnote{\url{https://pytorch.org/}} and our code can be found online.\footnote{\url{https://github.com/shayangharib/AUDASC}}
\vspace{-6pt}
\subsection{Dataset and data preprocessing}
\vspace{-6pt}
The dataset used for the development and evaluation of our method is the one provided as the development dataset of Task 1, subtask B, of the DCASE 2018 challenge~\cite{dcase2018:arxiv}. The dataset is collected with three different recording devices. The main recording device which is referred to as device A consists of a binaural microphone and a recorder using 48 kHz sampling rate and 24 bit resolution. The data from this device were re-sampled and averaged into a single channel to match the characteristic of data recorded by device B and C. The rest of data have been recorded using customer devices such as smart phones and cameras which are referred to as device B and C. This dataset contains a total of 28 hours of audio out of which 24 hours are from device A, 2 hours from device B, and 2 hours from device C. The proposed evaluation setup by the organizers of the DCASE 2018 challenge, 30$\%$ of audio files of device A, and 25$\%$ of device B and C are dedicated to the validation set. 

During the development of our method, the annotations of evaluation/test data of the Task 1, subtask B, were not publicly available. Therefore, we use the original (i.e. proposed by the evaluation setup of the DCASE Task 1, subtask B) validation data as our test data (referred to as test data for the rest of this paper), a randomly selected 10$\%$ of the original training data as our validation (referred to as validation data from now on), and the rest of the original training data as our actually training data (referred to as training data from now on). This means that we use 5510 files from device A, 486 files from device B, and 486 files from device C as training data. 612, 54, and 54 files from device A, B, and C, respectively, are used as our validation set. We test our method on 2518, 180, and 180 files, from devices A, B, and C respectively which is equivalent to original validation set of the Task 1, subtask B. 

From the available files, we extracted 64 log Mel-band energies, using a 2048 samples ($\sim 46$ ms) Hamming window and 50$\%$ overlap. The extracted features from the data recorded from device A is our source domain data. The rest (B and C) are our target domain data. We use the Librosa package for feature extraction.\footnote{\url{https://librosa.github.io/librosa/}} Since the amount of data for the target domain (i.e. the data from the B and C devices) are less than the source domain data (i.e. the data from device A), we oversampled the data from target domain to have the same amount of training data as the number of samples from device A, approximately 5.6 times more than the original size. 
\begin{figure*}[!ht]
\centering
\subfloat[
Confusion matrix for the non-adapted Kaggle model
]{%
\includegraphics[trim={.25cm, .4cm, .25cm, .5cm},clip,width=.9\columnwidth]{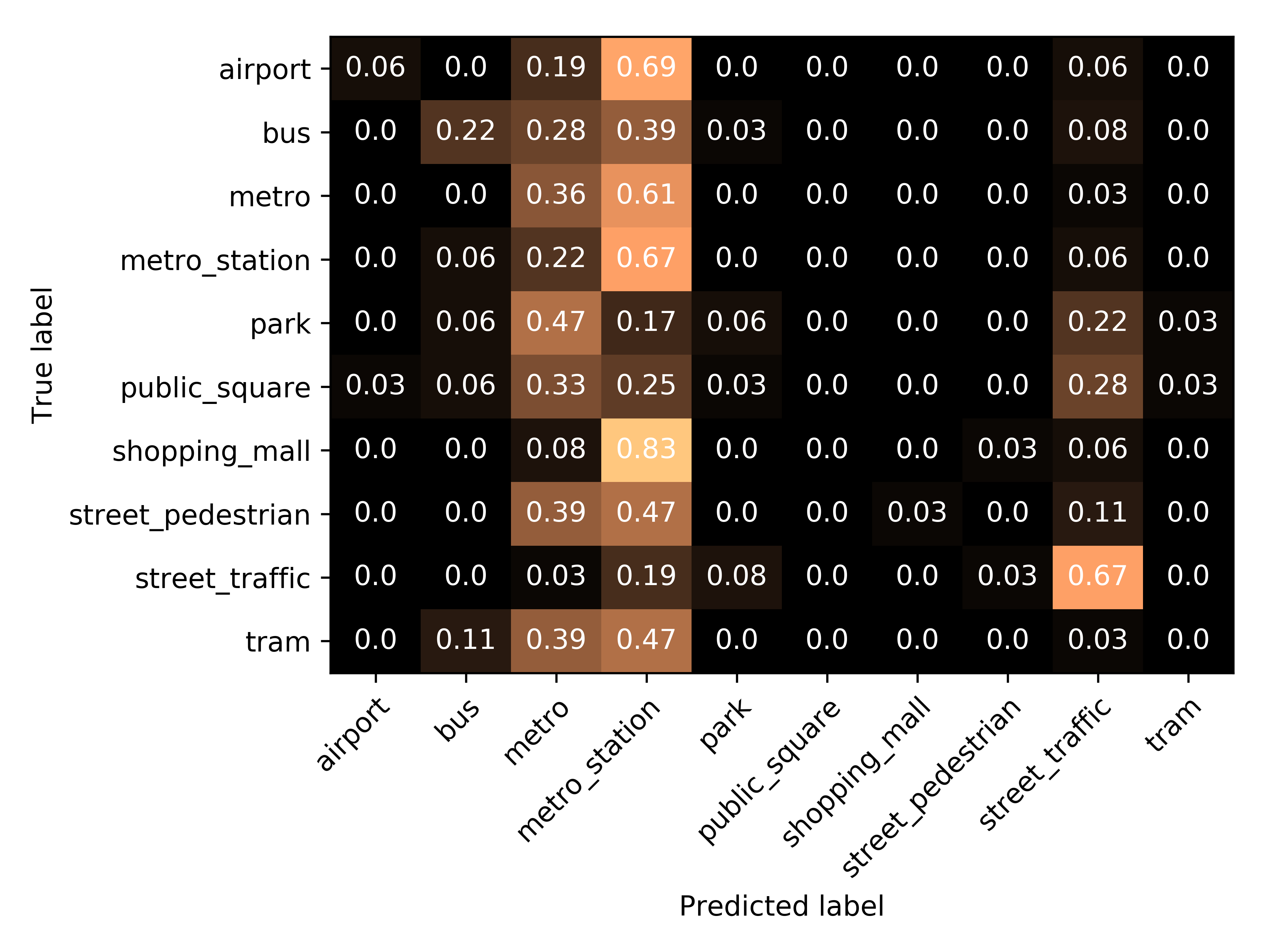}
\label{fig:conf-matr-non-adapted-target}
}
\subfloat[
Confusion matrix for the adapted Kaggle model
]{%
\includegraphics[trim={.25cm, .4cm, .25cm, .5cm},clip,width=.9\columnwidth]{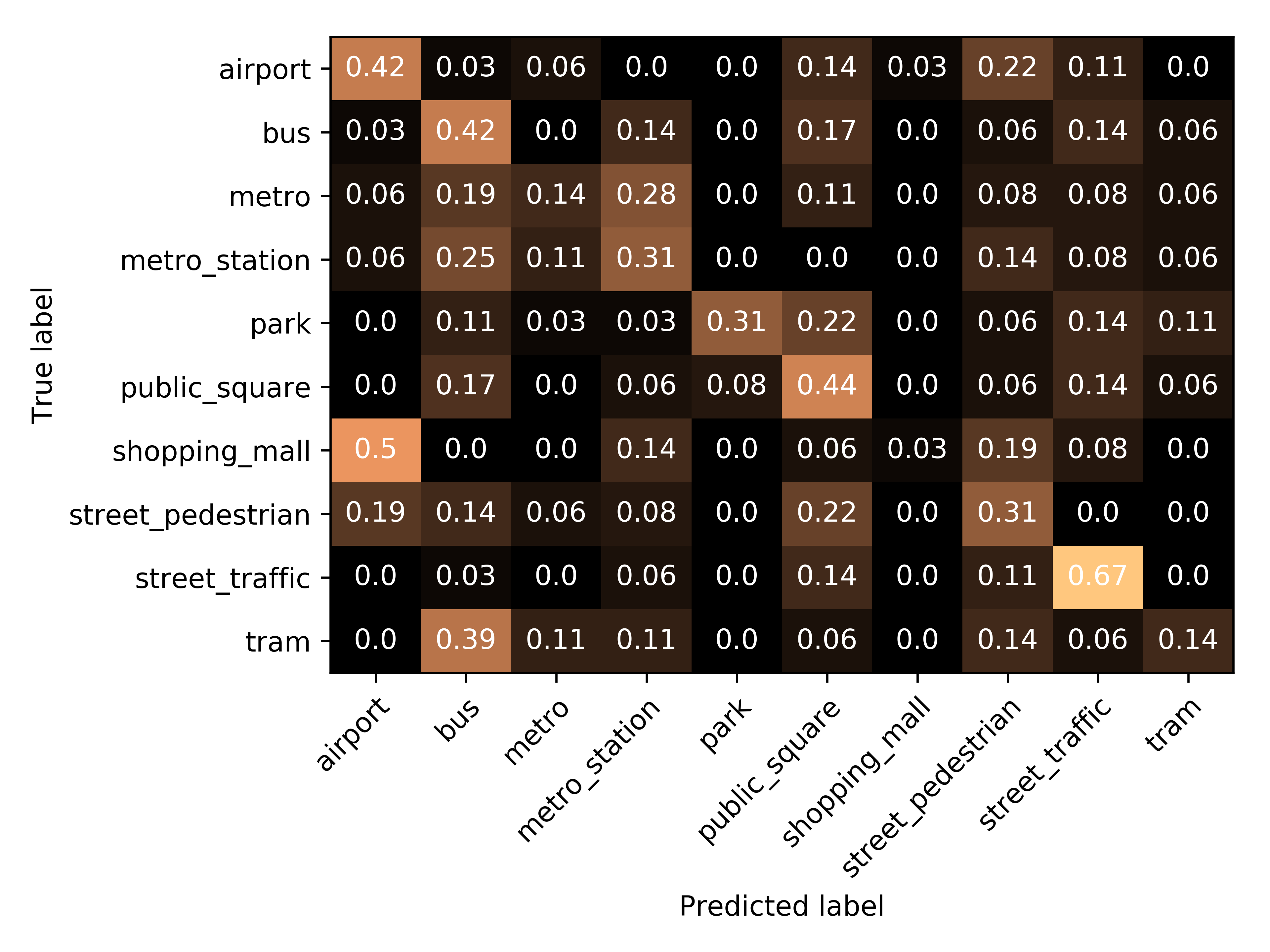}
\label{fig:conf-matr-adapted-target}
}
\caption{Confusion matrix of the \textbf{non-adapted}, (a), and \textbf{adapted}, (b), \textbf{Kaggle model} for the \textbf{target domain}. The values are normalized according to the amount of examples in each class. Brighter color indicates higher value.}
\label{fig:confmatrix}
\end{figure*}
\vspace{-8pt}
\subsection{Models used}
\vspace{-6pt}
Since our proposed method is independent of the employed model, we evaluate it on two different and published models. The first is the Kaggle model and the second is DCASE baseline model. The Kaggle model is mainly a convolutional neural network (CNN) which has 5 convolutional layers, with kernel sizes of \{(11, 11), (5, 5), (3, 3), (3, 3), (3,3)\} and amount of channels/filters of \{48, 128, 192, 192, 128\}. The first two convolutional layers use strides of (2,3) and the rest (1,1). The first two convolutional layers together with the last one are followed by rectified linear unit (ReLU) non-linearity, max pooling layer, and batch normalization. The rest convolutional layers are followed only by the ReLU non-linearity. DCASE model consists of two convolutional layers with 32 and 64 filters, respectively. Both layers have a (7, 7) kernel size followed by batch normalization, ReLU non-linearity, and a max pooling operation. The kernels of the pooling operations are \{(5, 5), (4, 100)\}. 

We use 64 log mel-band energies, but for the development of the DCASE model, 40 mel band energies were used. Therefore, we had to slightly alter the DCASE model in order to utilize our data. Specifically, we altered the kernel of the first pooling operation and the padding of the second convolutional layer. That is, we used kernel size of (8,4) for the pooling operation and we specified the padding for the second convolutional layer at (3,0). Because the Kaggle and the DCASE models had a different dimensionality of their outputs, we used two different discriminators. 

As the discriminator $D$, when employing the Kaggle model, we use three convolutional layers all with a kernel size of (3,3) and \{64, 32, 16\} as the number of channels/filters. All layers are followed by the ReLU non-linearity and batch normalization. The output of the third convolutional layer is flattened and given as an input to a linear layer, which outputs the prediction for samples as source or target. As a discriminator for the DCASE model we used one linear layer. The input to our label classifier $C$ is the output of the last layer of the model $M$ which is turned to a vector (i.e. flattened) and is given as an input to three for the Kaggle and two for the DCASE model linear layers followed by 25$\%$, for the Kaggle model, and 30$\%$, for the DCASE model, dropout. The non-linearity of all except the last layer of the classifier is the ReLU. Lastly, the output of our classifier is followed by a softmax non-linearity. 
\vspace{-8pt}
\subsection{Training and testing procedure}
\vspace{-6pt}
For the pre-training step, we use a minibatch size of 38 samples, all selected from source domain. During the domain adaptation process we used a minibatch size of 16 samples, out of which 10 were selected from the source domain and 6 from the target domain (more specifically 3 from device B and 3 from device C). For the pre-training and the domain adaptation process, Adam was selected as the optimizer with learning rate of $1\mathrm{e}{-4}$ and other values according to the ones presented in the original paper~\cite{adam}. We updated the parameters of the $M_{S}$ and $M_{T}$ after each iteration but (according to experimental observations) we updated the parameters of the discriminator $D$ after 10 iterations. We stopped the optimization procedure in pre-training and domain adaptation processes after 350 and 300 epochs respectively.
\vspace{-8pt} 
\section{Results and discussion}\label{sec:results}
\vspace{-6pt} 
We report the obtained accuracy of the label classification when using the non-adapted models (i.e. in the pre-training step) and when using the models after the domain adaptation process (i.e. adapted models), using the data from source and target domains. Table~\ref{tab:results1} presents the obtained accuracy on the source and target domains when using the Kaggle model and when using the DCASE model, respectively.
\vspace{-6pt} 
\begin{table}[!ht]
\caption{Obtained accuracy for the non-adapted and adapted Kaggle and DCASE models.}
\label{tab:results1}
\centering
\smallskip
\begin{tabular}{p{.6cm}cc|cc}
 & \multicolumn{2}{c}{Kaggle model} & \multicolumn{2}{c}{DCASE model}\\
\textit{} & \textbf{Non adapted} & \textbf{Adapted} & \textbf{Non adapted} & \textbf{Adapted}\\
\hline
Source & 65.25$\%$ & 65.37$\%$ & 61.71$\%$ & 61.23$\%$ \\
Target & 20.28$\%$ & 31.67$\%$ & 19.17$\%$ & 25.28$\%$ \\
\end{tabular}
\end{table}

Additionally, we present the confusion matrices of the label classification for the target domain of the non-adapted and adapted Kaggle model in Figure~\ref{fig:confmatrix}, where can be seen that the adapted model manages to increase significantly the correctly classified examples from all labels. This is easily visualized by the diagonal of the confusion matrices, where in Figures~\ref{fig:conf-matr-non-adapted-target} and~\ref{fig:conf-matr-adapted-target} there is a considerable difference. 
Additionally, our proposed method manages to increase the performance of the classification for different models. That is, no matter the architecture of the model $M$, by following our proposed method there is an increase on the target domain without significant decrease in the performance for the source domain. In fact, we managed to increase also the performance on the source domain for the adapted model. This is apparent in Table~\ref{tab:results1}, where the obtained accuracy for the adapted models and the target domain is greater, compared to the non-adapted. Furthermore, from the same table can be seen that the reduction in the accuracy at the source domain is around $0.5\%$ for the DCASE model. The accuracy is marginally (i.e. $\sim0.1\%$) greater for the source domain and the adapted model. We attribute this small increase to the usage of the label classifier $C$ during the domain adaptation process. 
\vspace{-8pt}
\section{Conclusions and future work}\label{sec:conclusions}
\vspace{-6pt}
We presented the first unsupervised adversarial domain adaptation for acoustic scene classification, which is also independent of the actual models used. The goal of our method is the adaptation of a pre-trained model on a source dataset, to a new and unseen target dataset. In a GAN-like setting, the adapting model tries to fool a discriminator that its output comes from the source dataset, while the non-adapted model informs the discriminator about the data that really coming from the source dataset. 

We managed to increase the performance of the used models to the unseen dataset by approx. $10\%$. This indicates that the domain adaption approaches can provide an appealing solution for the problem of mismatched training and testing data, regarding the acoustic scene classification. As future directions we suggest the adoption of different GAN losses and the usage of domain adaptation for sound event detection. 
\vspace{-8pt}
\section{ACKNOWLEDGMENT}\label{sec:ack}
\vspace{-6pt}
S. Gharib, E. \c{C}akir, K. Drossos, and T. Virtanen wish to acknowledge the CSC-IT Center for Science, Finland, for computational resources. Part of the computations leading to these results was performed on a TITAN-X GPU donated by NVIDIA to K. Drossos. Part of the research leading to these results has received funding from the European Research Council under the European Union’s H2020 Framework Programme through ERC Grant Agreement 637422 EVERYSOUND.
\bibliographystyle{IEEEtran}
\bibliography{refs}
\end{sloppy}
\end{document}